\documentclass[aps,reprint,floatfix,citeautoscript,longbibliography,superscriptaddress]{revtex4-1}
\usepackage{graphicx}
\usepackage{bm,amsmath,amssymb,mathrsfs,dcolumn}
\usepackage{txfonts}
\usepackage{gensymb}
\usepackage{epstopdf}
\usepackage{epsfig}
\usepackage[usenames]{color}
\usepackage{subfigure}
\usepackage{ulem}
\usepackage{float}
\usepackage{soul}
\usepackage{xcolor}
\usepackage{cancel}
\usepackage{hyperref}
\usepackage[marginal]{footmisc}
\hypersetup{colorlinks=true, linkcolor=blue, citecolor=blue, urlcolor=blue}
\setstcolor{red}

\setcitestyle{open={[},close={]},citesep={,\!},numbers}
\begin{document}

\preprint{APS/123-QED}

\title{Explosive growth of bistability in a cavity magnonic system}
\author{Meng-Xia Bi}
\affiliation{School of Science, Xi'an University of Posts and Telecommunications, Xi'an 710121, China}
\author{Huawei Fan}
\email{huawei.fan@xupt.edu.cn}
\affiliation{School of Science, Xi'an University of Posts and Telecommunications, Xi'an 710121, China}
\author{Wenting Wu}
\affiliation{North West Electric Power Design Institute, Xi'an, 710075, China}
\author{Jing-Jing He}
\affiliation{College of Information Science and Technology $\&$ Artificial Intelligence, Nanjing Forestry University, Nanjing 210037, China}
\author{Ming-Liang Hu}
\email{mingliang0301@163.com}
\affiliation{School of Science, Xi'an University of Posts and Telecommunications, Xi'an 710121, China}
\author{Xiao-Hong Yan}
\email{yanxh@njupt.edu.cn}
\affiliation{College of Science, Nanjing University of Posts and Telecommunications, Nanjing 210046, China}


\date{\today}

\begin{abstract}
  We conduct a theoretical investigation into explosive growth of bistability in a cavity magnonic system incorporating magnetic nonlinearity. In this system, the coupling between the magnon and photon generates the cavity magnon polaritons. When driving the photon-like polariton mode, the bistability can undergo a sudden transition with the increase of the driving power, resulting in an explosive growth of the bistable region by several times. Conversely, driving the magnon-like polariton mode only gives rise to normal bistability. This depends on whether the minimum driving power required to generate the bistability is non-monotonic with respect to the driving frequency. In addition, despite driving only the photon-like polariton mode, the photon- and magnon-like polariton modes can show simultaneous explosive growth of the bistability in microwave transmission, owing to the light-matter interaction. Our research sheds light on the hidden side of the nonlinear cavity magnonic system and provides a potential application for cavity spintronic devices founded on this novel feature.
\end{abstract}

\pacs{Valid PACS appear here}

\maketitle

\section{\label{sec:level1}Introduction}
Hybrid systems~\cite{WVEB:2006,XAYN:2013,AKM:2014,BGGW:2021}, by combining the advantages of different physical systems, can reveal more fascinating and unique physical effects that are difficult to observe in a single system, making them play an important role in uncovering the natural laws and developing new technologies. As one of the hybrid systems developed in the past decade, cavity magnonic system~\cite{LTGUN:2019,BK:2019,YCKDY:2022,ZVHULNHTBB:2022}, in which the magnon (the quanta of collective spin excitations~\cite{B:1930, HP:1940, D:1956}) in the yttrium iron garnet (YIG) crystal with low dissipation rate ($\sim$$1$ MHz) and high spin density ($\sim$$4.22\times10^{27} $m$^{-3}$) coherently~\cite{SF:2010,HZLHGMGG:2013,TIIYUN:2014,ZZJT:2014,GFCFKT:2014,LHF:2015,CYHGB:2015,BHCFXH:2015,ZWLLWNY:2015} or dissipatively~\cite{GSX:2018,HYYYRGSH:2018,BKJKYCK:2019,WRYXGYYH:2019,YWYX:2019,YYZHXY:2020,YWRGYLH:2020,ZFXSY:2023} couples to the microwave photon, has become a focus of extensive research because of its enormous potential in the exploration of fundamental physics and applications in information processing. Many interesting phenomena have been demonstrated by both theoretical and experimental studies, such as level repulsion~\cite{HZLHGMGG:2013,TIIYUN:2014,ZZJT:2014}, magnon dark modes~\cite{ZZZMJT:2015,HHBMH:2016}, exceptional points~\cite{ZLWLY:2017,ZY:2019,CY:2019} and surfaces~\cite{ZDZXJ:2019,GX:2022}, level attraction~\cite{HYYYRGSH:2018,BKJKYCK:2019}, nonreciprocity~\cite{WRYXGYYH:2019,WXXZY:2022}, Floquet ultrastrong coupling~\cite{XZHJJZ:2020}, coherent microwave amplification~\cite{YGRZLH:2023}, long-distance strong coupling~\cite{RWYCZL:2023,YYXFLH:2024}, and so on.

Benefiting from the rich magnetic properties in YIG crystals, recently the magnon Kerr effect induced by the magnetocrystalline anisotropy has been effectively excited under the strong driving~\cite{WZZLXWLHY:2016}, extending the research of the cavity magnonics from the linear field to the nonlinear field. The introduction of the Kerr effect significantly distorts the physical image, resulting in the folding behavior of the system, such as bistability~\cite{WZZLHY:2018,HYGZYH:2018,ZWY:2019,KXW:2019} where the system has two coexisting stable states. The direction of the hysteresis loop of the bistability is influenced by the positive and negative Kerr effects that depend on the angle between the external magnetic field and the crystallographic axis~\cite{WZZLHY:2018}. To achieve the bistability at low power, some non-Hermitian cavity magnonic systems have been designed and fabricated to enhance the Kerr nonlinearity, thereby reducing the threshold of the bistability~\cite{NMA:2021J,NMA:2021M,PYAH:2022,ZWX:2023}. For the YIG crystal exhibiting magnetostriction, which mediates the coupling between magnetization and elastic strain of the magnetic material, mechanical bistability arises through the utilization of the Kerr effect~\cite{SLFWY:2022}. Furthermore, when the system involves multiple YIG crystals or engages with other nonlinear mechanisms, in addition to bistability, multistability can also be realized by adjusting the system parameters reasonably~\cite{NZSA:2020,BYZX:2021,SWLZAY:2021,BFCHY:2024}. Very recently, a special form of nonlinearity termed hidden multistability has been demonstrated in this hybrid system~\cite{BFYL:2024}, where the intermediate steady state is folded into the bistable hysteresis loop. These nonlinear studies make the cavity magnonic system a promising platform for the development of devices such as switches, memories, and logic gates.

The interaction between the magnon and photon generates a quasiparticle known as the cavity magnon polaritons (CMPs)~\cite{HZLHGMGG:2013,TIIYUN:2014,ZZJT:2014} that is a mixture of the magnon and photon. The characteristics of the CMPs depend on the mixing ratio of the two. For the case where the magnon component is greater than the photon component, the CMPs can be described as the magnon-like polariton mode. Conversely, it can be described as the photon-like polariton mode. Previous nonlinear researches have mainly focused on driving the magnon-like polariton mode~\cite{WZZLHY:2018,HYGZYH:2018,ZWY:2019}, with little investigation into driving the photon-like polariton mode. This inevitably leads to the possibility of missing some important physical phenomena. In this paper, we study the nonlinear behavior of the cavity magnon system with the Kerr effect via driving the photon-like polariton mode. Indeed, very different from the previous studies~\cite{WZZLHY:2018,HYGZYH:2018,ZWY:2019}, a sudden transition of the bistability happens in such case, causing an explosive growth in bistable region. This interesting phenomenon can be explained based on the minimum driving power required to generate the bistability. The critical magnon frequency of this sudden transition is influenced by the system parameter. Moreover, the photon- and magnon-like polariton modes display simultaneous explosive growth of the bistability in the transmission spectrum of the cavity. Our results contribute to enhancing the understanding of nonlinear dynamics in cavity magnonic systems.

The remainder of this paper is organized as follows. In Sec.~\ref{sec:level2}, we construct the model of the cavity magnonic system and derive the equation of motion, while also obtaining the transmission coefficient of the cavity. Based on the formula derived in Sec.~\ref{sec:level2}, we demonstrate the explosive growth of the bistability and the impact of the system parameter on it in Sec.~\ref{sec:level3}. Also, the nonlinear transmission spectrum of the cavity is investigated in this section. Finally, discussion and conclusion are given in in Sec.~\ref{sec:level4}.

\section{\label{sec:level2}Model and Theory}

\subsection{\label{sec:level01}Effective Hamiltonian}
\begin{figure}[htbp]  
  \centering
  \includegraphics[width=0.45\textwidth]{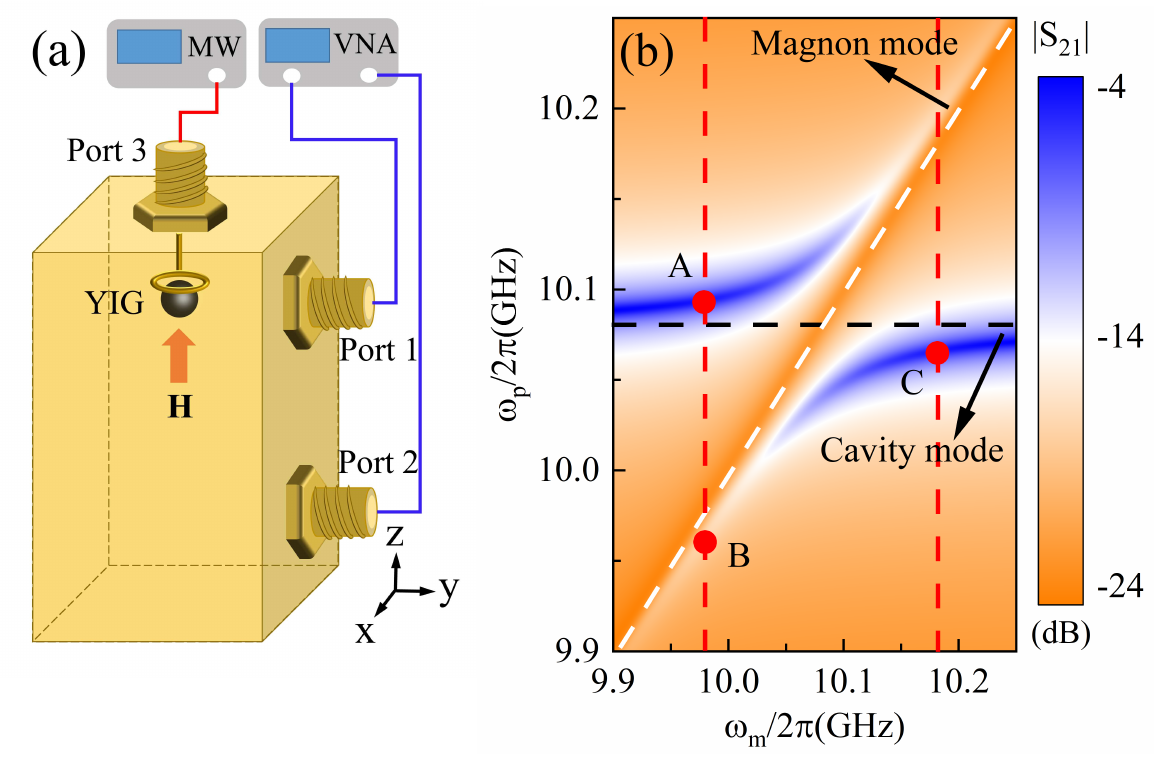}\\
  \caption{(a) Schematic layout of the cavity magnonic system. The uniformly biased magnetic field $\mathbf{H}$ applied to the YIG sphere is for aligning the magnetization and tuning the magnon frequency. Ports 1 and 2 are for measuring the transmission spectrum of the cavity by using a VNA. Port 3 is for directly driving the YIG sphere by using a loop antenna connected to an MV. (b) Transmission spectrum of the CMPs versus the magnon frequency $\omega_{m}/2\pi$ and the probe-field frequency $\omega_{p}/2\pi$ at a low driving power $P_{d}=0.01$ mW. The black and white dashed lines represent the uncoupled cavity and magnon modes, respectively. The two vertical red dashed lines, respectively, correspond to $\omega_{m}/2\pi=9.98$ and $10.18$ GHz at which we show the nonlinear behavior of the hybrid system.}\label{f1}
\end{figure}

The hybrid system we considered is a YIG sphere of diameter 1 mm coupled to a 3D microwave cavity with inner dimensions 44.0$\times$22.0$\times$6.0 mm$^{3}$~\cite{WZZLHY:2018}, as schematically shown in Fig.~\ref{f1}(a), where a uniformly biased magnetic field $\mathbf{H}$ is applied to the YIG sphere to align the magnetization and tune the magnon frequency. The saturation magnetization and surface roughness of the YIG sample are 143.2 kA/m and 0.1 $\mu$m, respectively~\cite{SWLZAY:2021}. Ports 1 and 2, connected to a vector network analyzer (VNA), are used to probe the transmission spectrum of the cavity. A loop antenna, connected to a microwave (MW) source, is used to directly drive the YIG sphere through port 3. In our study, we concentrate on the coupling between the cavity mode TE$_{102}$ and the Kittel mode with all spins uniformly precessing in phase under $\mathbf{H}$~\cite{WZZLHY:2018}.

Based on the model mentioned above, the effective Hamiltonian of the hybrid system consists of the following five parts:
\begin{align}\label{e1}
    H&=H_{c}+H_{m}+H_{I}+H_{d}+H_{p},
\end{align}
where $H_{c}=\omega_{c}a^{\dag}a$ (in units of $\hbar$) is the bare Hamiltonian of the photon, with the creation (annihilation) operator $a^{\dag}$ ($a$) at frequency $\omega_{c}$. Denoting $\mathbf{H}=H\mathbf{e}_{z}$, the YIG sphere with the volume $V_{m}$ has a Hamiltonian:
\begin{align}\label{e2}
   H_{m}&=-\mu_{0}\int_{V_{m}}\textbf{M}\cdot\textbf{H}d\rho-\frac{\mu_{0}}{2}\int_{V_{m}}\textbf{M}\cdot\textbf{H}_{an}d\rho,
\end{align}
where the two terms on the right-hand side of the equation, respectively, are the Zeeman energy and the magnetocrystalline anisotropic energy, $\mu_{0}$ and $\textbf{M}=(M_{x}, M_{y}, M_{z})$, respectively, are the vacuum magnetic permeability and magnetization of the YIG sphere, and $\textbf{H}_{an}$ is the magnetocrystalline anisotropic field which can be expressed as~\cite{M:1951,WZZLXWLHY:2016}
\begin{align}\label{e3}
   \textbf{H}_{an}&=-\frac{\upsilon_{x}K_{an}M_{x}}{\mu_{0}M^{2}_{s}}\textbf{e}_{x}-\frac{\upsilon_{y}K_{an}M_{y}}{\mu_{0}M^{2}_{s}}\textbf{e}_{y}-
   \frac{\upsilon_{z}K_{an}M_{z}}{\mu_{0}M^{2}_{s}}\textbf{e}_{z},
\end{align}
where $\upsilon_{x (y, z)}$ is a dimensionless coefficient that depends on the direction of the crystallographic axis along the biased magnetic field, $K_{an}$ is the dominant first-order anisotropy constant, $M_{s}$ is the saturation magnetization, and $\textbf{e}_{x (y, z)}$ is the unit vector in the $x (y, z)$ direction. Substituting Eq.~(\ref{e3}) into Eq.~(\ref{e2}) and then using the relation~\cite{SF:2010} $\textbf{M}=-\gamma\textbf{S}/V_{m}$ where $\gamma$ and $\textbf{S}$ are the gyromagnetic ratio and macrospin operator, respectively, we have
\begin{align}\label{e4}
   H_{m}&=\gamma\mu_{0}HS_{z}+\frac{\gamma^{2}K_{an}}{2M^{2}_{s}V_{m}} \left(\upsilon_{x}S_{x}^{2}+\upsilon_{y}S_{y}^{2}+\upsilon_{z}S_{z}^{2}\right).
\end{align}
Here, the macrospin operator is defined as $\textbf{S}$=$\sum_{j}\textbf{s}_{j}$=($S_{x}$, $S_{y}$, $S_{z}$), where $\textbf{s}_{j}$ represents the spin operator of the $j$th spin in the YIG sphere and $\sum_{j}$ denotes the summation over all spins within the YIG sphere. Inserting the raising and lowing operators of the macrospin $S^{\pm}=S_{x}\pm iS_{y}$ into Eq.~(\ref{e4}), it can be rewritten as
\begin{align}\label{e5}
   H_{m}&=\gamma\mu_{0}HS_{z}+\frac{K_{x}}{4}\left(S^{+}+S^{-}\right)^{2}-\frac{K_{y}}{4}\left(S^{+}-S^{-}\right)^{2}\nonumber \\
   &\quad+K_{z}S_{z}^{2}
\end{align}
with $K_{x (y, z)}=\upsilon_{x (y, z)}\gamma^{2}K_{an}/2M^{2}_{s}V_{m}$.

For the YIG sphere embedded in the microwave cavity, the spins couple to the cavity photon via the magnetic-dipole interaction:
\begin{align}\label{e6}
   H_{I}&=g_{s}\left(S^{+}+S^{-}\right)\left(a^{\dag}+a\right),
\end{align}
where $g_{s}$ is the coupling strength between each spin and the photon. When the YIG sphere is directly driven by the MV, the driving Hamiltonian can be described as
\begin{align}\label{e7}
   H_{d}&=\Omega_{d}\left(S^{+}+S^{-}\right)\left(e^{-i\omega_{d} t}+e^{i\omega_{d} t}\right),
\end{align}
where $\Omega_{d}$ and $\omega_{d}$ are the driving strength and driving frequency, respectively. The transmission spectrum of the cavity is measured by the VNA, giving rise to the probing Hamiltonian:
\begin{align}\label{e8}
   H_{p}&=\eta\left(a^{\dag}+a\right)\left(e^{-i\omega_{p} t}+e^{i\omega_{p} t}\right),
\end{align}
where $\eta$ and $\omega_{p}$ are the probe-field strength and probe-field frequency,  respectively. Combining the bare Hamiltonian of the photon and Eqs.~(\ref{e5})-(\ref{e8}) and then using the rotating-wave approximation~\cite{WM:1994}, we obtain the total effective Hamiltonian of the hybrid system:
\begin{align}\label{e9}
   H&=\omega_{c}a^{\dag}a+\gamma\mu_{0}HS_{z}+\frac{K_{x}+K_{y}}{4}\left(S^{+}S^{-}+S^{-}S^{+}\right)+K_{z} S_{z}^{2}\nonumber \\
   &\quad+g_{s}\left(a^{\dag}S^{-}+aS^{+} \right)+\Omega_{d}\left(S^{+}e^{-i\omega_{d} t}+S^{-}e^{i\omega_{d} t}\right)\nonumber\\
    &\quad+\eta\left(a^{\dag}e^{-i\omega_{p} t}+ae^{i\omega_{p} t}\right).
\end{align}

\subsection{\label{sec:level02}Equation of motion and transmission coefficient}
Applying the master equation method~\cite{BP:2007} to the Hamiltonian in Eq.~(\ref{e9}), the equations of motion for the expectation values of the operator $\langle a\rangle $, $\langle S^{-}\rangle$, and $\langle S_{z} \rangle$ can be written as (see the detail in Appendix A)
\begin{align}\label{e10}
\begin{split}
     \langle \dot{a}\rangle&=-i\left[\left(\omega_{c}-i\frac{\kappa_{c}}{2}\right)\langle a\rangle+g_{s}\langle S^{-}\rangle+\eta e^{-i\omega_{p}t}\right],\\
    \langle \dot{S}^{-}\rangle&= -i\left[\left(\gamma\mu_{0}H-i\frac{\gamma_{m}}{2}\right)\langle S^{-}\rangle+2K_{m}\langle S_{z}\rangle\langle S^{-}\rangle\right.\\
       &\left. \quad -2g_{s}\langle S_{z}\rangle\langle a\rangle-2\Omega_{d} \langle S_{z}\rangle e^{-i\omega_{d}t} \right],\\
   \langle \dot{S}_{z}\rangle&= -i\left[-i\frac{\gamma_{m}}{2}\left(\langle S_{z}\rangle+\frac{N}{2} \right)+g_{s}\left( \langle S^{+}\rangle\langle a\rangle-\langle a^{\dag}\rangle\langle S^{-}\rangle\right)\right.\\
       &\left. \quad+\Omega_{d}\left(\langle S^{+}\rangle e^{-i\omega_{d}t}-\langle S^{-}\rangle e^{i\omega_{d}t} \right)  \right],
\end{split}
\end{align}
where $\kappa_{c}$ and $\gamma_{m}$, respectively, are the photon and magnon dissipation rates, $K_{m}=\left(2K_{z}-K_{x}-K_{y}\right)/2$ is the magnon Kerr coefficient, and $N$ is the total spin number in the YIG sphere. For the hybrid system subject to the strong driving, the numbers of the photon and magnon are considerable. Therefore, we write the operator in the form of the expectation value in Eq.~(\ref{e10}). For convenience, we introduce the transformation of the variables, i.e., $\langle a\rangle=\sqrt{N}\alpha$, $\langle S^{-}\rangle=NS$, $\langle S^{+}\rangle=NS^{\ast}$, and $\langle S_{z}\rangle=NS^{z}$, into Eq.~(\ref{e10}), leading to
\begin{align}\label{e11}
\begin{split}
      \dot{\alpha}&=-i\left[\left(\omega_{c}-i\frac{\kappa_{c}}{2}\right)\alpha+GS+\xi e^{-i\omega_{p}t}\right],\\
      \dot{S}&= -i\left[\left(\gamma\mu_{0}H-i\frac{\gamma_{m}}{2}\right)S+2K S^{z}S-2G S^{z}\alpha \right.\\
             &\left. \quad -2\Omega_{d}  S^{z} e^{-i\omega_{d}t} \right],\\
      \dot{S}^{z}&= -i\left[-i\frac{\gamma_{m}}{2}\left( S^{z}+\frac{1}{2} \right)+G\left(  S^{\ast}\alpha-S\alpha^{\ast}\right)  \right.\\
                 &\left. \quad   +\Omega_{d}\left(S^{\ast} e^{-i\omega_{d}t}-S e^{i\omega_{d}t} \right)  \right],
\end{split}
\end{align}
where $G=g_{s}\sqrt{N}$ is the photon-magnon coupling strength, $K=K_{m}N$ and $\xi=\eta/\sqrt{N}$. Since the driving field is much stronger than the probe field, the expectation value of the operator can be divided into the following two parts:
\begin{equation}\label{e12}
 \alpha= A_{d}e^{-i\omega_{d}t}+A_{p}e^{-i\omega_{p}t}, \quad
 S= S_{d}e^{-i\omega_{d}t}+S_{p}e^{-i\omega_{p}t},
\end{equation}
where $A_{d}$ ($A_{p}$) and $S_{d}$ ($S_{p}$) are the amplitudes due to the driving (probe) field. Due to the weak probe field, $A_{p}$ and $S_{p}$ are treated as the perturbation. It is note that we have neglected other frequency components of the perturbation in Eq.~(\ref{e12}), as the focus in our study is on the transmission coefficient of the probe field. Substituting Eq.~(\ref{e12}) into Eq.~(\ref{e11}), we can obtain two sets of equations of motion. One is about driving:
\begin{align}\label{e13}
\begin{split}
      \dot{A}_{d}&=-i\left[\left(\omega_{c}-\omega_{d}-i\frac{\kappa_{c}}{2}\right)A_{d}+GS_{d}\right],\\
  \dot{S}_{d}&= -i\left[\left(\omega_{m}+\Delta_{m}-\omega_{d}-i\frac{\gamma_{m}}{2}\right)S_{d}-2G S^{z}A_{d}-2\Omega_{d} S^{z} \right],\\
   \dot{S}^{z}&= -i\left[-i\frac{\gamma_{m}}{2}\left( S^{z}+\frac{1}{2} \right)+G\left(  S^{\ast}_{d}A_{d}-S_{d}A^{\ast}_{d}\right)+\Omega_{d}\left(S^{\ast}_{d}-S_{d}\right)\right],
\end{split}
\end{align}
and the other is about probing:
\begin{align}\label{e14}
\begin{split}
      \dot{A}_{p}&=-i\left[\left(\omega_{c}-\omega_{p}-i\frac{\kappa_{c}}{2}\right)A_{p}+GS_{p}+\xi\right],\\
  \dot{S}_{p}&= -i\left[\left(\omega_{m}+\Delta_{m}-\omega_{p}-i\frac{\gamma_{m}}{2}\right)S_{p}-2G S^{z}A_{p} \right],
\end{split}
\end{align}
where $\omega_{m}=\gamma\mu_{0}H-K$ is the magnon frequency, $\Delta_{m}=K(1+2S^{z})$ is the magnon frequency shift due to the Kerr effect. Here we only consider the first-order terms of $A_{p}$ and $S_{p}$, and neglect their higher-order terms in Eq.~(\ref{e14}). Equation~(\ref{e13}) is our starting point for numerically studying the nonlinear behavior of the system in this paper. Using the steady-state condition for $A_{d}$, $S_{d}$, $A_{p}$, and $S_{p}$ (i.e., $\dot{A}_{d}=0$, $\dot{S}_{d}=0$, $\dot{A}_{p}=0$, and $\dot{S}_{p}=0$), we have
\begin{align}\label{e15}
\begin{split}
     & \left(\delta_{c}-i\frac{\kappa_{c}}{2}\right)A_{d}+GS_{d}=0,\\
     & \left(\delta_{m}+\Delta_{m}-i\frac{\gamma_{m}}{2}\right)S_{d}-2G S^{z}A_{d}-2\Omega_{d} S^{z}=0,\\
     & -i\frac{\gamma_{m}}{2}\left( S^{z}+\frac{1}{2} \right)+G\left(  S^{\ast}_{d}A_{d}-S_{d}A^{\ast}_{d}\right)+\Omega_{d}\left(S^{\ast}_{d}-S_{d}\right)=0,
\end{split}
\end{align}
and
\begin{align}\label{e16}
\begin{split}
      &\left(\delta_{c1}-i\frac{\kappa_{c}}{2}\right)A_{p}+GS_{p}+\xi=0,\\
      &\left(\delta_{m1}+\Delta_{m}-i\frac{\gamma_{m}}{2}\right)S_{p}-2G S^{z}A_{p}=0,
\end{split}
\end{align}
where $\delta_{c (m)}=\omega_{c (m)}-\omega_{d}$ is the photon (magnon) frequency detuning relative to the driving frequency, and $\delta_{c1 (m1)}=\omega_{c (m)}-\omega_{p}$ is the photon (magnon) frequency detuning relative to the probe-field frequency. By solving Eq.~(\ref{e15}), we can obtain a cubic equation of $\Delta_{m}$:
\begin{align}\label{e17}
   &D_{3}\Delta_{m}^{3}+D_{2}\Delta_{m}^{2}+D_{1}\Delta_{m}+D_{0}=0,
\end{align}
with
\begin{align}\label{e18}
\begin{split}
      &D_{3}=\left(K+\frac{\delta_{c}G^{2}}{\delta_{c}^{2}+\frac{\kappa_{c}^{2}}{4}}\right)^{2}
            +\left(\frac{\frac{\kappa_{c}}{2}G^{2}}{\delta_{c}^{2}+\frac{\kappa_{c}^{2}}{4}}\right)^{2},\\
      &D_{2}=2K\left[\left(\delta_{m}-\frac{\delta_{c}G^{2}}{\delta_{c}^{2}+\frac{\kappa_{c}^{2}}{4}}\right)
             \left(K+\frac{\delta_{c}G^{2}}{\delta_{c}^{2}+\frac{\kappa_{c}^{2}}{4}}\right)\right.\\
             &\left. \qquad -\frac{\frac{\kappa_{c}}{2}G^{2}}{\delta_{c}^{2}+\frac{\kappa_{c}^{2}}{4}}
             \left(\frac{\gamma_{m}}{2}+\frac{\frac{\kappa_{c}}{2}G^{2}}{\delta_{c}^{2}+\frac{\kappa_{c}^{2}}{4}}\right) \right],\\
      &D_{1}=K^{2}\left[\left(\delta_{m}-\frac{\delta_{c}G^{2}}{\delta_{c}^{2}+\frac{\kappa_{c}^{2}}{4}}\right)^{2}
             +\left(\frac{\gamma_{m}}{2}+\frac{\frac{\kappa_{c}}{2}G^{2}}{\delta_{c}^{2}+\frac{\kappa_{c}^{2}}{4}}\right)^{2}+4\Omega_{d}^{2}\right],\\
      &D_{0}=-4K^{3}\Omega_{d}^{2}.
\end{split}
\end{align}
Since the driving strength $\Omega_{d}$ cannot be directly measured in the experiment, it can be converted into a relationship with the driving power $P_{d}$, i.e., $2K\Omega_{d}^{2}=cP_{d}$ in which $c$ is an experimental fitting coefficient characterizing the coupling strength between the driving field and magnon~\cite{WZZLHY:2018}. For the theoretical expression of $\Omega_{d}$, it can be rigorously derived in Ref.~\cite{LZA:2018}. Equation~(\ref{e17}) is our starting point for analytically studying the nonlinear behavior of the system in this paper. By solving Eq.~(\ref{e16}), we can obtain
\begin{align}\label{e19}
   A_{p}&=\frac{-\xi}{\delta_{c1}-i\frac{\kappa_{c}}{2}+\frac{2G^{2}S^{z}}{\delta_{m1}+\Delta_{m}-i\frac{\gamma_{m}}{2}}}.
\end{align}
Using the input-output theory~\cite{WM:1994}, the probe fields at the input port 1 and output port 2 can be written, respectively, as
\begin{align}\label{e20}
   a_{in}^{p}=\frac{\xi}{\sqrt{\kappa_{c1}}},\quad a_{out}^{p}=i\sqrt{\kappa_{c2}}A_{p},
\end{align}
where $\kappa_{c1}$ and $\kappa_{c2}$ are the external dissipation rates of the photon due to the input port 1 and output port 2, respectively. Combining Eqs.~(\ref{e19}) and (\ref{e20}), the transmission coefficient of the cavity can be expressed as
\begin{align}\label{e21}
  S_{21}=\frac{a_{out}^{p}}{a_{in}^{p}}=\frac{\sqrt{\kappa_{c1}\kappa_{c2}}}{i\delta_{c1}+\frac{\kappa_{c}}{2}-\frac{2G^{2}S^{z}}
  {i\left(\delta_{m1}+\Delta_{m}\right)+\frac{\gamma_{m}}{2}}}.
\end{align}
Equation~(\ref{e21}) involves the magnon frequency shift $\Delta_{m}$ induced by the Kerr effect, reflecting the nonlinearity of the microwave transmission.

Prior to exploring the strong driving, the linear case under the weak driving is presented as a basic image. Using Eq.~(\ref{e21}), we plot the transmission spectrum versus the magnon frequency $\omega_{m}/2\pi$ and the probe-field frequency $\omega_{p}/2\pi$ at a low driving power of $P_{d}=0.01$ mW, as shown in Fig.~\ref{f1}(b). Because of the strong coherent coupling between the photon and magnon, a clear and typical level repulsion emerges, featuring lower and upper branches (i.e., the CMPs), when the magnon mode intersects the cavity mode. However, since the Kerr effect is very weak under low excitation, no nonlinearity is observed in Fig.~\ref{f1}(b). For this calculation, the parameters, derived from the experimental reports~\cite{WZZLHY:2018,HYGZYH:2018,SWLZAY:2021}, are set as follows: $\omega_{c}/2\pi=10.08$ GHz, $\omega_{d}/2\pi=10.12$ GHz, $G/2\pi=40$ MHz, $\kappa_{c}/2\pi=3.8$ MHz, $\kappa_{c1}/2\pi=\kappa_{c2}/2\pi=1.0$ MHz, $\gamma_{m}/2\pi=16.5$ MHz, $K/2\pi=3.0$ GHz, and $c/(2\pi)^{3}=10.0$ MHz$^{3}$/mW.
According to the relationship between the magnon frequency and the external magnetic field in Eq.~(\ref{e14}), the range of the external magnetic field in Fig.~\ref{f1}(b) is from 460 mT to 475 mT, which implies that the magnetization of the YIG sphere is close to saturation.
The low drive power in Fig.~\ref{f1}(b) is compared to the high drive power required to induce a significant magnon frequency shift. The magnon frequency shift that is less than half of the magnon dissipation rate (i.e., $\Delta_{m}<\gamma_{m}/2$) can be considered as indicating a weak nonlinear effect, corresponding to the driving power $P_{d}$ less than 20 mW. In such cases, the system can be approximated linearly.

\section{\label{sec:level3}Results}

\subsection{\label{sec:level03}Explosive growth of bistability}
\begin{figure}[htbp]  
  \centering
  \includegraphics[width=0.45\textwidth]{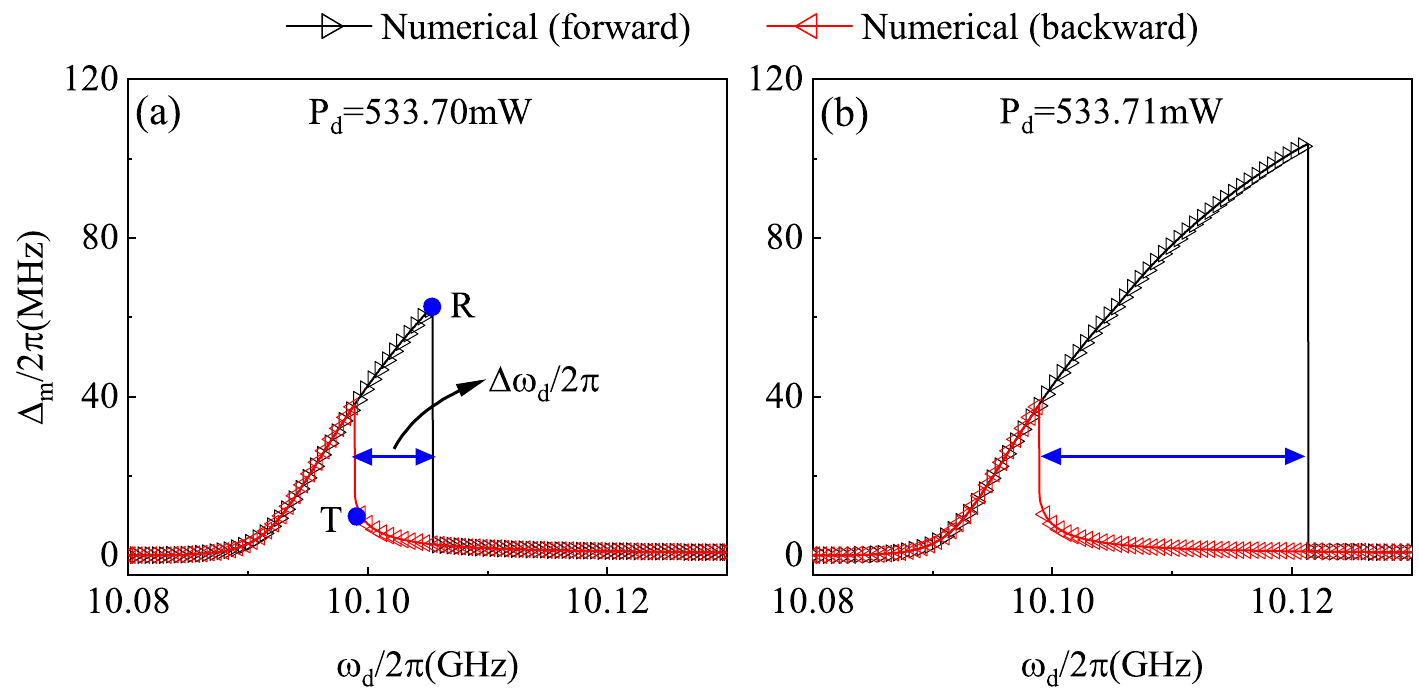}\\
  \caption{Numerically simulated magnon frequency shift $\Delta_{m}/2\pi$ versus the driving frequency $\omega_{d}/2\pi$ for (a) $P_{d}=533.70$ mW and (b) $P_{d}=533.71$ mW when the photon-like polariton mode is driven at $\omega_{m}/2\pi=9.98$ GHz. The black and red triangle curves represent the numerically forward and backward sweeps of $\omega_{d}/2\pi$, respectively.}\label{f2}
\end{figure}

With the increase of the driving power, the YIG sphere is strongly driven and generates considerable magnons, which amplifies the Kerr nonlinear effect and therefore can dramatically modify the dynamic behavior of the system. When the magnon frequency $\omega_{m}/2\pi$ is tuned to be $9.98$ GHz [corresponding to the left vertical red dashed line in Fig.~\ref{f1}(b)], we plot the magnon frequency shift $\Delta_{m}/2\pi$ versus the driving frequency $\omega_{d}/2\pi$ for different values of the driving power $P_{d}$ by using the numerical method, as shown in Fig.~\ref{f2}. Here the numerical method is used to simulate the experiment by numerically solving Eq.~(\ref{e13}), in which $\omega_{d}/2\pi$ is adiabatically increased or decreased with a step $\delta \omega_{d}/2\pi=0.05$ MHz. In Fig~\ref{f2}, the driving frequency remains above the cavity photon frequency (i.e., $\omega_{d}/2\pi>\omega_{c}/2\pi$), indicating that the upper branch is driven. For $\omega_{m}/2\pi=9.98$ GHz (which deviates to the left from the resonance point), the upper branch [i.e., point $A$ in Fig.~\ref{f1}(b)] is the photon-like polariton mode due to its higher photon component. In this case, as a result of the Kerr nonlinearity, the magnon frequency shift $\Delta_{m}/2\pi$ displays bistability by numerically sweeping $\omega_{d}/2\pi$ forward (the black triangle curve) and backward (the red triangle curve)  with $P_{d}=533.70$ mW, as shown in Fig.~\ref{f2}(a). Point $R$ ($T$) represents the switching point of $\Delta_{m}/2\pi$ for the forward (backward) sweep, where the up (down) state suddenly switches to the down (up) state. The difference in $\omega_{d}/2\pi$ between these two points is described as the bistable region $\Delta\omega_{d}/2\pi$, in which the up and down states coexist. However, when the driving power is slightly increased from $533.70$ mW to $533.71$ mW in Fig.~\ref{f2}(b) (here, a step size of 0.01 mW is adopted as the minute increment for simulating the continuous variation of the driving power), a sudden discontinuous transition occurs in the bistability, with the bistable region $\Delta\omega_{d}/2\pi$ growing explosively to about $3.5$ times as that shown in Fig.~\ref{f2}(a). This amazing property offers a mean for the rapid adjustment of the bistability.

\begin{figure}[htbp]  
  \centering
  \includegraphics[width=0.45\textwidth]{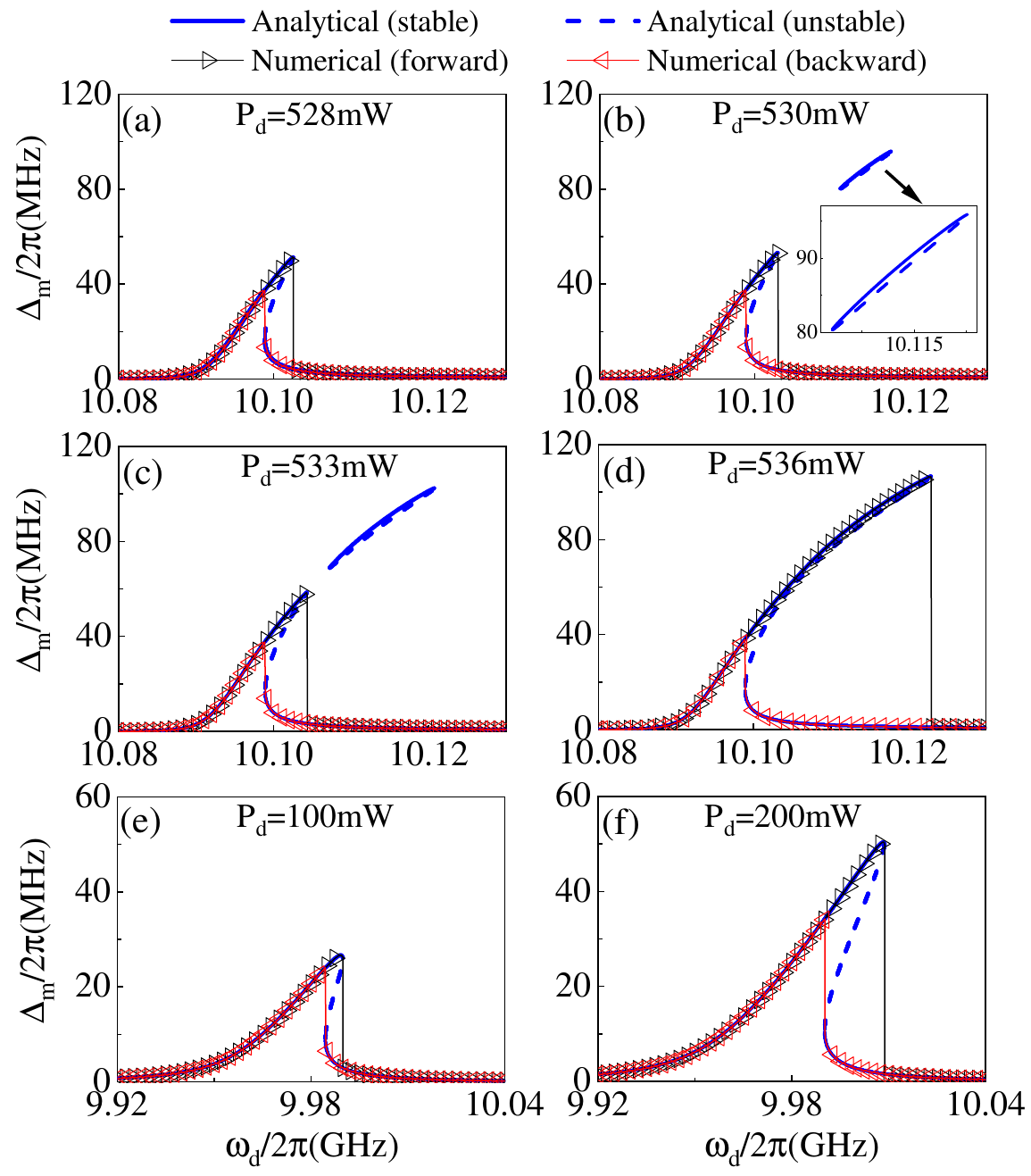}\\
  \caption{Analytically calculated and numerically simulated magnon frequency shift $\Delta_{m}/2\pi$ versus the driving frequency $\omega_{d}/2\pi$ for different values of the driving power $P_{d}$ when $\omega_{m}/2\pi=9.98$ GHz. (a)-(d) are for driving the photon-like polariton mode. The inset in (b) is the magnified view of the isolated branch. (e) and (f) are for driving the magnon-like polariton mode. The blue solid and dashed lines denote the analytically stable and unstable states, respectively. The black and red triangle curves represent the numerically forward and backward sweeps of $\omega_{d}/2\pi$, respectively.}\label{f3}
\end{figure}

To further observe the formation process of the sudden transition, we calculate dependence of the magnon frequency shift $\Delta_{m}/2\pi$ on the driving frequency $\omega_{d}/2\pi$ under different values of the driving power $P_{d}$, and present the analytical results in addition to the numerical ones, as shown in Figs.~\ref{f3}(a)--\ref{f3}(d). The analytical results, as determined by Eq.~(\ref{e17}), are denoted by the blue solid (stable state) and dashed (unstable state) lines, respectively. For $P_{d}=528$ mW, the magnon frequency shift $\Delta_{m}/2\pi$ tilts to the right and produces the normal bistability shown in Fig.~\ref{f3}(a). The numerical simulation agrees well with the analytical steady state. However, as $P_{d}$ increases to $530$ mW, an intriguing phenomenon emerges: apart from the main part, there is also an analytical isolated branch in Fig.~\ref{f3}(b), which is noticeably different from that of the previous studies~\cite{WZZLHY:2018,HYGZYH:2018,ZWY:2019}. Surprisingly, although this isolated branch is stable [see the blue solid line in the inset of Fig.~\ref{f3}(b)], it cannot be directly observed via conventional numerically forward and backward sweeps of the driving frequency $\omega_{d}/2\pi$, making it experimentally undetectable. Therefore, the numerical simulation still exhibits a small bistable region. When $P_{d}$ reaches $533$ mW in Fig.~\ref{f3}(c), the isolated branch becomes larger and closer to the main part, while the bistable region slightly expands in the numerical result. As $P_{d}$ further increases, e.g., $P_{d}=536$ mW, the isolated branch eventually merges with the main part, forming a unified structure, as shown in Fig.~\ref{f3}(d). It can be seen that due to the mergence of the isolated branch and main part, the structure of the bistability undergoes a sudden transition from Fig.~\ref{f3}(c) to Fig.~\ref{f3}(d), thus resulting in an explosive growth of the bistable region in the simulated experiment.

When the driving frequency is controlled within the range less than the cavity photon frequency, i.e., $\omega_{d}/2\pi < \omega_{c}/2\pi$, the lower branch is driven, as shown in Figs.~\ref{f3}(e) and~\ref{f3}(f). Since it contains more magnon component for $\omega_{m}/2\pi=9.98$ GHz, the lower branch [i.e., point $B$ in Fig.~\ref{f1}(b)] is the magnon-like polariton mode. In this situation, with the increase of the driving power, the bistable region continuously increases without any sudden transitions, thus leading to the normal bistability. Compared to Fig.~\ref{f3}(a), the lower driving power in Fig.~\ref{f3}(f) results in a larger bistable region. It is because the Kerr nonlinearity originates from the magnon mode, and the magnon component in the magnon-like polariton mode is more abundant compared to that in the photon-like polariton mode, thus making it easier for the magnon-like polariton mode to exhibit the bistability. This is consistent with previous research~\cite{BYXD:2020}.

\subsection{\label{sec:level04}Theoretical explanation}
In Fig.~\ref{f3}(b), the emergence of the isolated branch is due to the fact that the critical driving power threshold required to trigger bistability in this region is lower than that at the discontinuity. As the driving power continues to increase beyond the bistability excitation threshold at the discontinuity, bistability gradually appears at the discontinuity, accompanied by the expansion of the isolated branch, as described by Fig.~\ref{f3}(c). Eventually, the isolated branch merges with the main body in Fig.~\ref{f3}(d), resulting in the explosive growth of the bistability. To further prove this point, we rewrite Eq.~(\ref{e17}) as follows:
\begin{align}\label{e22}
  D_{3}\Delta_{m}^{3}+D_{2}\Delta_{m}^{2}+  D_{1}^{\prime} \Delta_{m}-2cKP_{d}\left(K-\Delta_{m}\right)&=0,
\end{align}
with
\begin{align}\label{e23}
  D_{1}^{\prime}&=K^{2}\left[\left(\delta_{m}-\frac{\delta_{c}G^{2}}{\delta_{c}^{2}+\frac{\kappa_{c}^{2}}{4}}\right)^{2}
             +\left(\frac{\gamma_{m}}{2}+\frac{\frac{\kappa_{c}}{2}G^{2}}{\delta_{c}^{2}+\frac{\kappa_{c}^{2}}{4}}\right)^{2}\right],
\end{align}
where $\Omega_{d}$ is substituted with $P_{d}$ based on the relationship between $\Omega_{d}$ and $P_{d}$ given below Eq.~(\ref{e18}). Using the condition $dP_{d}/d\Delta_{m}=0$, i.e.,
\begin{align}\label{e24}
2D_{3}\Delta_{m}^{3}+\left(D_{2}-3KD_{3}\right)\Delta_{m}^{2}-2KD_{2}\Delta_{m}-K  D_{1}^{\prime} &=0,
\end{align}
we can obtain the switching points of the bistability [i.e., the root $\Delta_{m}$ of Eq.~(\ref{e24})] in the parameter space of $P_{d}$. Notably, among the three roots of $\Delta_{m}$ in Eq.~(\ref{e24}), only two roots of $\Delta_{m}$ can yield a positive driving power upon substitution into Eq.~(\ref{e22}). When inserting the lager root $\Delta_{m}$ into Eq.~(\ref{e22}), the minimum driving power $P_{d}^{\min}$ required to generate the bistability for a given $\omega_{d}/2\pi$ can be determined.

\begin{figure}[htbp]  
  \centering
  \includegraphics[width=0.47\textwidth]{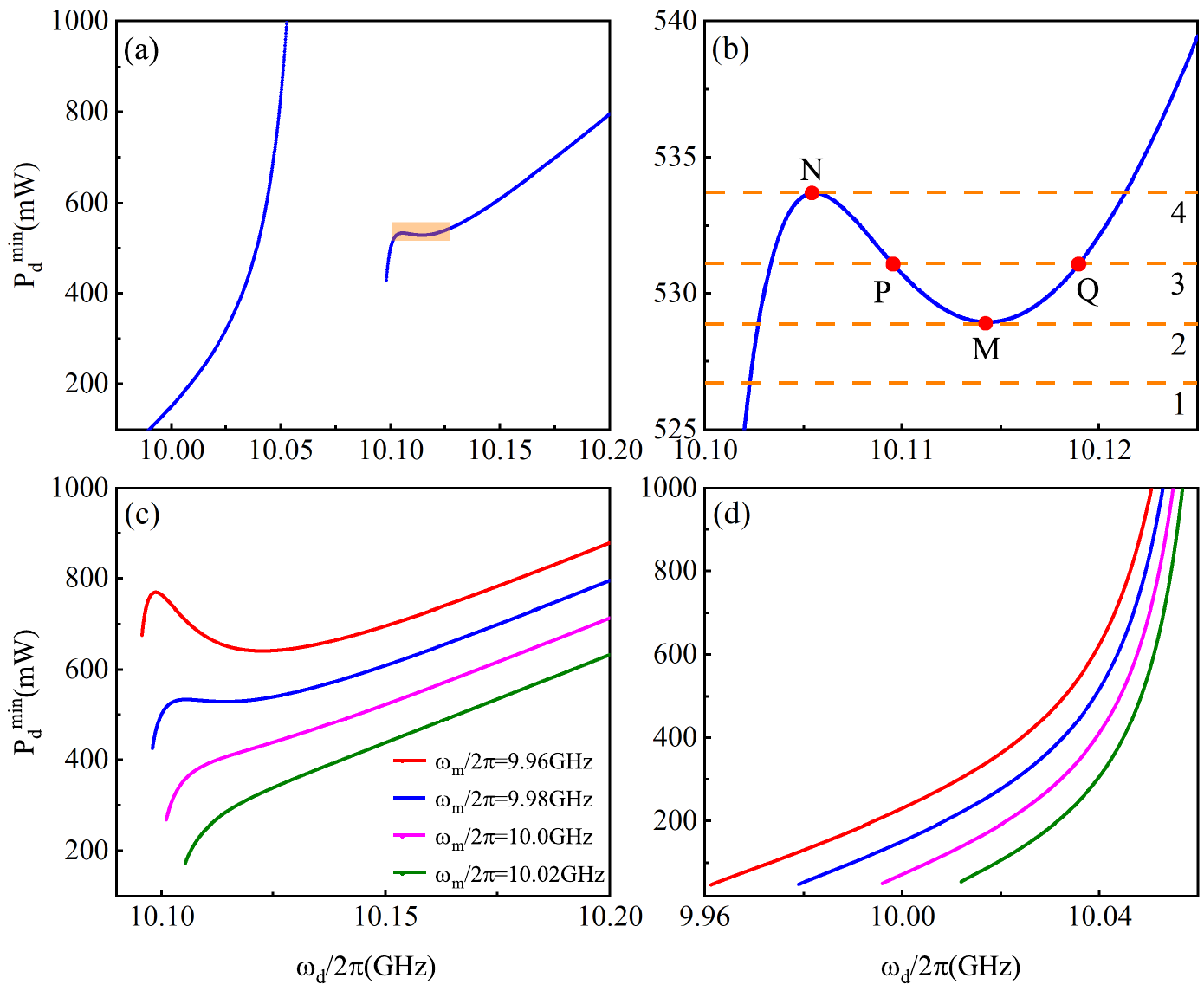}\\
  \caption{(a) Minimum driving power $P_{d}^{\min}$ versus the driving frequency $\omega_{d}/2\pi$ at $\omega_{m}/2\pi=9.98$ GHz. (b) The magnified view of the orange region in (a). The horizontal orange dashed lines 1-4 correspond to $P_{d}=526.82$, $528.94$, $531.22$, and $533.71$ mW, respectively. Points $M$ and $N$ are the extreme points, while points $P$ and $Q$ are the intersection points of the curve and the dashed line 3. (c) and (d) Minimum driving power $P_{d}^{\min}$ versus the driving frequency $\omega_{d}/2\pi$ for different values of the magnon frequency $\omega_{m}/2\pi$ when driving the upper and lower branches, respectively.}\label{f4}
\end{figure}

In Fig.~\ref{f4}(a), we plot the minimum driving power $P_{d}^{\min}$ versus the driving frequency $\omega_{d}/2\pi$ at $\omega_{m}/2\pi=9.98$ GHz. The left and right curves represent the driving of the lower and upper branches (i.e., the magnon- and photon-like polariton modes), respectively. The lower branch displays a monotonic behavior, indicating that the bistability gradually arises as the driving frequency $\omega_{d}/2\pi$ increases, thereby not producing the sudden transition. However, the upper branch exhibits a non-monotonic behavior. We plot the magnified view of the orange region marked in Fig.~\ref{f4}(a), as shown in Fig.~\ref{f4}(b). When $P_{d}$ is tuned to be $526.82$ mW, the bistability can be excited at $\omega_{d}/2\pi$ when $P_{d}^{\min}$ $<$ $526.82$ mW, i.e., the part where the blue curve lies below the dashed line $1$. Since this part is continuous, the bistability is uninterrupted, corresponding to, e.g., Fig.~\ref{f3}(a). For $P_{d}=528.94$ mW (i.e., dashed line 3), the part below this value starts to display a discontinuous behavior with a separated point $M$, indicating the onset of the isolated branch. As $P_{d}$ reaches $531.22$ mW (i.e., dashed line 3), the isolated branch expands to the region corresponding to the driving frequency from point $P$ to point $Q$, and tends to converge with the left part, as depicted in, e.g., Figs.~\ref{f3}(b) and~\ref{f3}(c). When $P_{d}$ increases to $533.71$ mW (i.e., dashed line 4), the isolated branch and the left part happen to converge at point $N$. As a result, $P_{d}^{\min}$ below $533.71$ mW begins to form a continuous whole, leading to the explosive growth of the bistable region, as illustrated in, e.g., Fig.~\ref{f2}(b). For $P_{d}>533.71$ mW, the system maintains the uninterrupted bistability, corresponding to, e.g., Fig.~\ref{f3}(d). Therefore, the non-monotonicity is responsible for the sudden transition of the bistability.

In Fig.~\ref{f4}(c), we plot the minimum driving power $P_{d}^{\min}$ versus the driving frequency $\omega_{d}/2\pi$ for different values of the magnon frequency $\omega_{m}/2\pi$ when driving the photon-like polariton mode. For a smaller magnon frequency, e.g., $\omega_{m}/2\pi=9.96$ GHz denoted by the red curve, due to the upward shift of the curve compared to the case of $\omega_{m}/2\pi=9.98$ GHz (i.e., the blue curve), the driving power $P_{d}$ required to achieve the sudden transition increases. As $\omega_{m}/2\pi$ increases, the curve progressively transitions from the non-monotonicity to monotonicity, resulting in the elimination of the extreme point and, consequently the disappearance of the sudden transition. However, when driving the magnon-like polariton mode, $P_{d}^{\min}$ versus $\omega_{d}/2\pi$ is always monotonic for different values of $\omega_{m}/2\pi$, indicating no sudden transition, as shown in Fig.~\ref{f4}(d).

\begin{figure}[htbp]  
  \centering
  \includegraphics[width=0.45\textwidth]{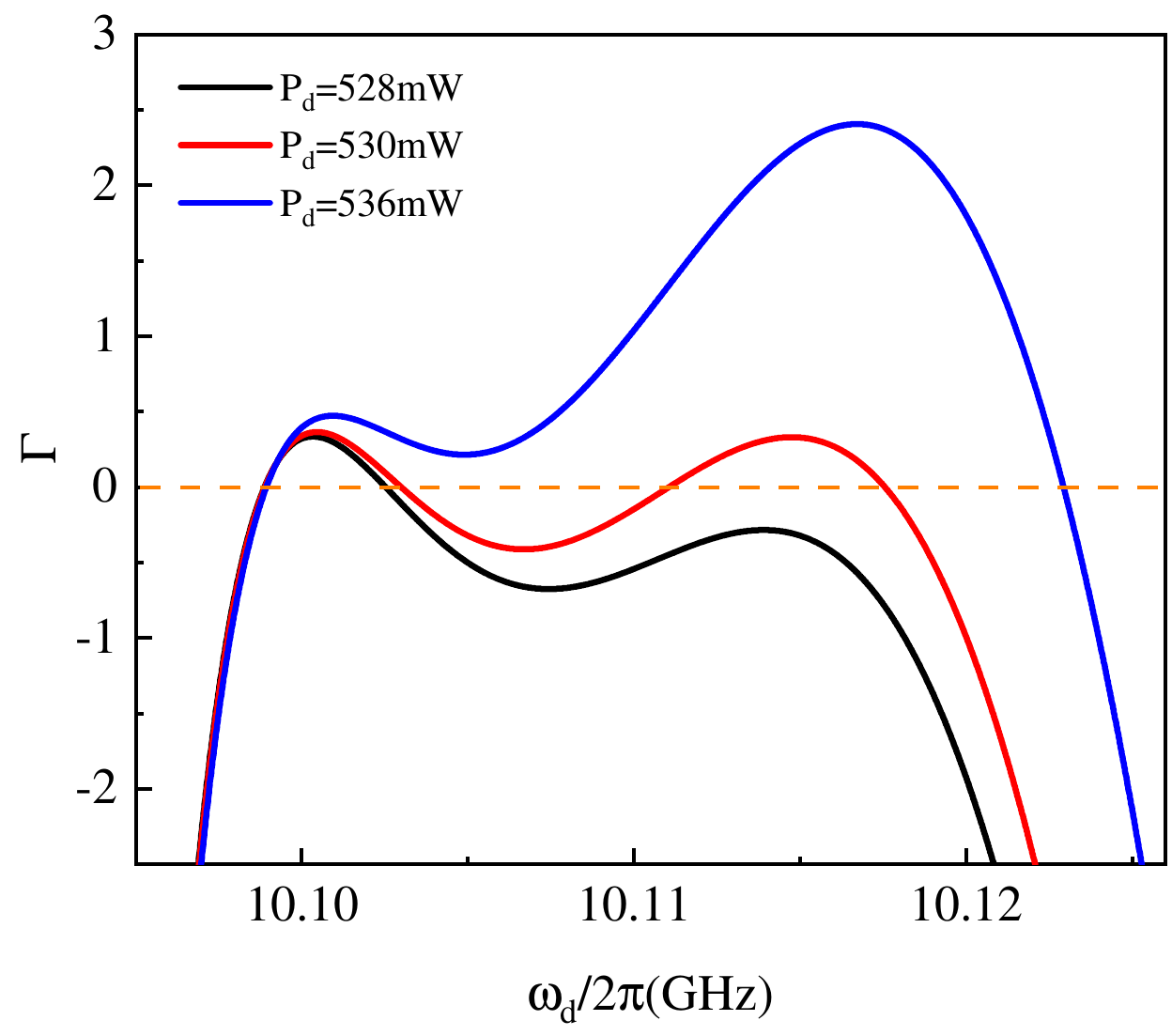}\\
  \caption{The discriminant $\Gamma$ of Eq.~(\ref{e17}) versus the driving frequency $\omega_{d}/2\pi$ for different values of the driving power $P_{d}$ when $\omega_{m}/2\pi=9.98$ GHz.}\label{f5}
\end{figure}

Essentially, the nonlinear behavior of the system is governed by the real roots of the cubic equation [i.e., Eq.~(\ref{e17})]. By plotting the discriminant $\Gamma$ of Eq.~(\ref{e17}) as a function of the driving frequency $\omega_{d}/2\pi$ in Fig.~\ref{f5} when $\omega_{m}/2\pi=9.98$ GHz, we can elucidate the origin of the explosive growth of bistability. For $P_{d}$=$528$ mW, the discriminant undergoes a transition from $\Gamma<0$ to $\Gamma>0$ and then back to $\Gamma<0$ (the black curve). In the region where $\Gamma<0$, equation~(\ref{e17}) has only one real root, corresponding to the monostable state in Fig.~\ref{f3}(a). For the region where $\Gamma>0$, equation~(\ref{e17}) has three real roots (two stable states and one unstable state), corresponding to the bistable region in Fig.~\ref{f3}(a). Thus, the entire process exhibits the normal bistability. However, when $P_{d}$=$530$ mW, the discriminant transitions from $\Gamma<0$ to $\Gamma>0$, then to $\Gamma<0$, followed by $\Gamma>0$, and finally back to $\Gamma<0$ (the red curve). Unlike the previous case, an additional $\Gamma>0$ region appears here, which corresponds to the isolated branch in Fig.~\ref{f3}(b). As the driving power continues to increase, for example, to $P_{d}$=$536$ mW, the $\Gamma>0$ region reduces to one again (the blue curve). This process signifies the merging of the isolated branch with the main part, as elaborated in Fig.~\ref{f3}(d). These transitions clarify that the isolated branch and the explosive growth of bistability are direct consequences of the discriminant's sign changing with the driving frequency, which could provide insights for the design of other systems.

\subsection{\label{sec:level05}Critical magnon frequency}
From the above analysis one can see that the sudden transition disappears as the magnon frequency $\omega_{m}/2\pi$ increases. To obtain the critical magnon frequency $\omega_{m}^{c}/2\pi$, $2cKP_{d}\left(K-\Delta_{m}\right)$ in Eq.~(\ref{e22}) can be approximated as $2cK^{2}P_{d}$ since $K/\Delta_{m}\approx 30$ in our study. So, equation~(\ref{e22}) can be simplified as
\begin{align}\label{e25}
  D_{3}\Delta_{m}^{3}+D_{2}\Delta_{m}^{2}+  D_{1}^{\prime} \Delta_{m}-2cK^{2}P_{d}&=0.
\end{align}
Using the condition $dP_{d}/d\Delta_{m}=0$, i.e.,
\begin{align}\label{e26}
  3D_{3}\Delta_{m}^{2}+2D_{2}\Delta_{m}+ D_{1}^{\prime}&=0,
\end{align}
we can obtain the switching points of the bistability in the parameter space of $P_{d}$ as follows
\begin{align}\label{e27}
  \Delta_{m\pm}&=\frac{-D_{2}\pm\sqrt{D_{2}^{2}-3D_{3}  D_{1}^{\prime} }}{3D_{3}}.
\end{align}
Inserting $\Delta_{m+}$ into Eq.~(\ref{e25}), the minimum driving power can be expressed as
\begin{align}\label{e28}
  P_{d}^{\min} &=\frac{2D_{2}^{3}-9D_{3}D_{2}  D_{1}^{\prime} -2(D_{2}^{2}-3D_{3}  D_{1}^{\prime} )^{\frac{3}{2}}}{54cK^{2}D_{3}^{2}}.
\end{align}
For $\Delta_{m-}$, it corresponds to the maximum power that produces bistability. Then, by solving $dP_{d}^{\min}/d\omega_{d}=0$, the driving frequency $\omega_{d}/2\pi$ corresponding to the extreme points $M$ and $N$ in Fig.~\ref{f4}(b) can be obtained for a given magnon frequency $\omega_{m}/2\pi$. We plot $\omega_{d}/2\pi$ at the extreme point versus $\omega_{m}/2\pi$, as depicted by the blue curve in Fig.~\ref{f5}(a). As $\omega_{m}/2\pi$ increases, the two driving frequencies $\omega_{d}/2\pi$ gradually converge, meet at $\omega_{m}/2\pi=9.98$ GHz marked by the black dot, and subsequently vanish. The black dot is the critical magnon frequency symbolled by $\omega_{m}^{c}/2\pi$ at which the sudden transition begins to occur. This is because the extreme points exist within the range where $\omega_{m}/2\pi$ is less than $\omega_{m}^{c}/2\pi$, but disappear when $\omega_{m}/2\pi$ exceeds $\omega_{m}^{c}/2\pi$.
\begin{figure}[htbp]  
  \centering
  \includegraphics[width=0.4\textwidth]{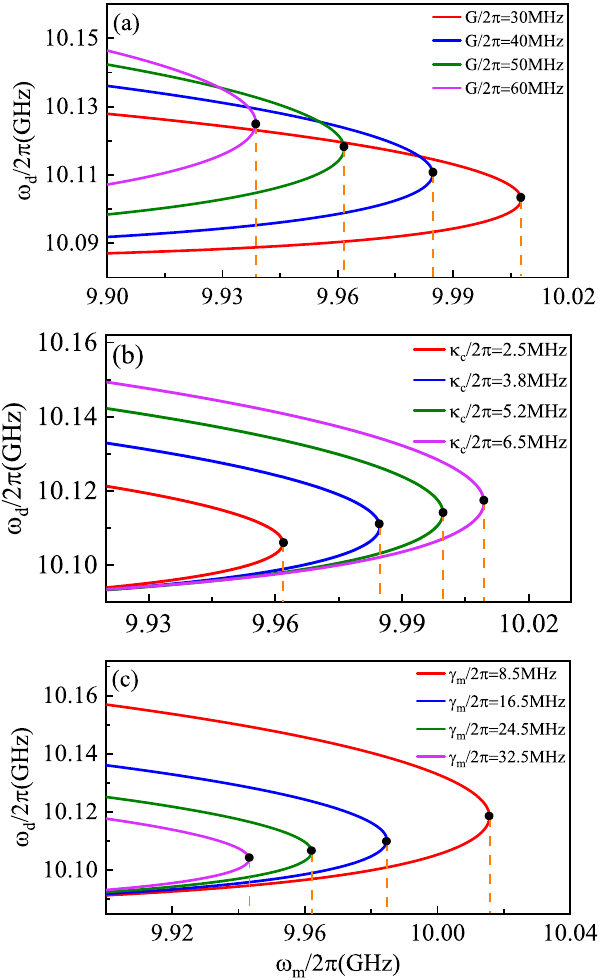}\\
  \caption{Driving frequency $\omega_{d}/2\pi$ at the extreme point versus the magnon frequency $\omega_{m}/2\pi$ for different values of (a) the coupling strength $G/2\pi$, (b) the photon dissipation rate $\kappa_{c}/2\pi$, and (c) the magnon dissipation rate $\gamma_{m}/2\pi$. The black dots denote the critical magnon frequencies $\omega_{m}^{c}/2\pi$ for different parameters.}\label{f6}
\end{figure}

Because of the controllability of the cavity magnonic system, the critical magnon frequency $\omega_{m}^{c}/2\pi$ can be adjusted by changing the system parameter, as shown in Fig.~\ref{f6}. We initially take into account the modulation effect exerted by the coupling strength $G/2\pi$ on $\omega_{m}^{c}/2\pi$. As $G/2\pi$ increases, $\omega_{m}^{c}/2\pi$ gradually decreases in Fig.~\ref{f6}(a). This is because for the same photon component in the upper branch, the greater the coupling strength, the smaller the corresponding magnon frequency. Secondly, the dissipation rate of the system also influences $\omega_{m}^{c}/2\pi$. As depicted in Fig.~\ref{f6}(b), increasing the photon dissipation rate $\kappa_{c}/2\pi$ leads to a corresponding increase in $\omega_{m}^{c}/2\pi$. The reason for this is that the minimum driving power rises with the increase of $\kappa_{c}/2\pi$, suggesting that the excitation of the nonlinear behavior becomes more difficult. Therefore, in order to generate the sudden transition, it is necessary for the system to increase $\omega_{m}^{c}/2\pi$ to reduce the photon component in the upper branch, thereby mitigating the impact due to the increase of $\kappa_{c}/2\pi$. Similarly, as the magnon dissipation rate $\gamma_{m}/2\pi$ increases, the system needs to decrease $\omega_{m}^{c}/2\pi$ to reduce the magnon component in the upper branch, thus mitigating the effect of the increased minimum driving power caused by the rise in $\gamma_{m}/2\pi$, as illustrated in Fig.~\ref{f6}(c).

\subsection{\label{sec:level06}Transmission spectrum}

\begin{figure}[htbp]  
  \centering
  \includegraphics[width=0.47\textwidth]{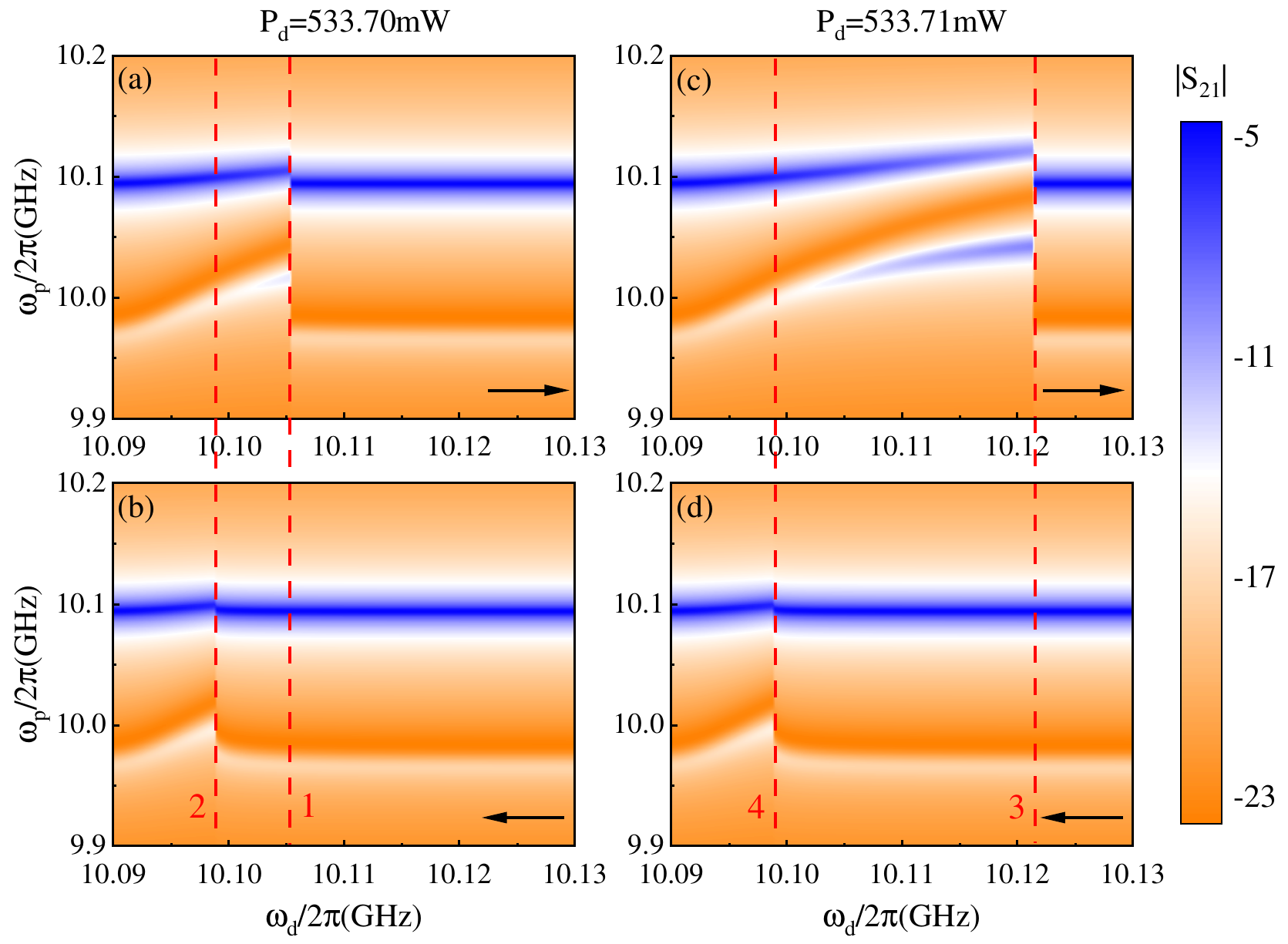}\\
  \caption{Transmission spectrum versus the driving frequency $\omega_{d}/2\pi$ and the probe-field frequency $\omega_{p}/2\pi$ under the high excitation when driving the photon-like polariton mode at $\omega_{m}/2\pi=9.98$ GHz. (a) and (b) represent, respectively, the forward and backward sweeps of $\omega_{d}/2\pi$ labeled by the black arrows when $P_{d} = 533.70$ mW, while (c) and (d) represent that when $P_{d}=533.71$ mW. The vertical red dashed lines 1-4 indicate the switching points of the bistability, respectively. Other parameters are the same as in Fig.~\ref{f2}.}\label{f7}
\end{figure}

We next investigate the explosive growth of the bistability from the perspective of the microwave transmission. Figure~\ref{f7} shows the two-dimensional contour of the transmission spectrum versus the driving frequency $\omega_{d}/2\pi$ and the probe-field frequency $\omega_{p}/2\pi$ under the high excitation when the photon-like polariton mode is driven at $\omega_{m}/2\pi = 9.98$ GHz. The transmission spectrum displays both the photon-like polariton mode (i.e., the upper branch) and the magnon-like polariton mode (i.e., the lower branch). When $P_{d}$ is set to $533.70$ mW and $\omega_{d}/2\pi$ is swept forward, a switching point of the polariton frequency occurs on both the upper and lower branches at the identical $\omega_{d}/2\pi$ indicated by the vertical red dashed line 1, as shown in Fig.~\ref{f7}(a). When sweeping $\omega_{d}/2\pi$ backward in Fig.~\ref{f7}(b), the switching points of the upper and lower branches appear at another identical $\omega_{d}/2\pi$ marked by the vertical red dashed line 2. Figures~\ref{f7}(a) and \ref{f7}(b) demonstrate a simultaneous bistability of both the photon- and magnon-like polariton modes, even though only the photon-like polariton mode is driven. This occurs because the Kerr effect induces a frequency shift of the magnon mode, which subsequently results in a frequency shift of the CMPs due to the coupling between the magnon and photon.
From a theoretical perspective, we can also elucidate the phenomenon of simultaneous bistability. By neglecting the dissipation terms in Eq.~(\ref{e16}), as well as the probe-field strength and probe-field frequency, we obtain the eigenvalues of the polariton modes generated by the coupling between the magnon and photon:
\begin{align}\label{e29}
  \lambda_{1,2}^{c}&=\frac{\omega_{c}+\omega_{m}+\Delta_{m}\pm\sqrt{\left(\omega_{c}-\omega_{m}-\Delta_{m}\right)^{2}+4G^{2}\left(1-\frac{\Delta_{m}}{K}\right)}}{2},
\end{align}
with $\lambda_{1}^{c}$ and $\lambda_{2}^{c}$ being the upper and lower branches, respectively. From Fig.~\ref{f2}(a), the driven photon-like polariton mode (i.e., the upper branch) induces the bistability of $\Delta_{m}$. Since both $\lambda_{1}^{c}$ and $\lambda_{2}^{c}$ contain $\Delta_{m}$, this bistability is simultaneously manifested in both branches. Consequently, the two polariton modes exhibit the simultaneous bistability in Figs.~\ref{f7}(a) and \ref{f7}(b). However, when the magnon and photon are decoupled (i.e., $G/2\pi=0$), Equation~(\ref{e29}) is reduced to the following form:
\begin{align}\label{e30}
  \lambda_{1}^{dc}=\omega_{c}, \quad \lambda_{2}^{dc}=\omega_{m}+\Delta_{m}.
\end{align}
It is not difficult to find that the polariton modes have degenerated into the pure photon and magnon modes. Since $\Delta_{m}$ exists solely within $\lambda_{1}^{dc}$, the simultaneous bistability does not occur in both $\lambda_{1}^{dc}$ and $\lambda_{2}^{dc}$.

With the driving power $P_{d}$ slightly increasing from $533.70$ mW to $533.71$ mW, however, the sweeping result of the transmission spectrum undergoes a notable transformation, wherein the bistable regions of the photon- and magnon-like polariton modes (marked by the vertical red dashed lines 3 and 4) simultaneously display the explosive growth, as shown in Figs.~\ref{f7}(c) and \ref{f7}(d). The reason for this is that the explosive growth of bistability of $\Delta_{m}$ in Fig.~\ref{f2}(b) is reflected in the two polariton modes [i.e., Eq.~(\ref{e29})]. This result suggests that the CMPs can serve as a bridge for the explosive growth of the bistability between different polariton modes.

\subsection{\label{sec:level07}Negative Kerr effect}
\begin{figure}[htbp]  
  \centering
  \includegraphics[width=0.48\textwidth]{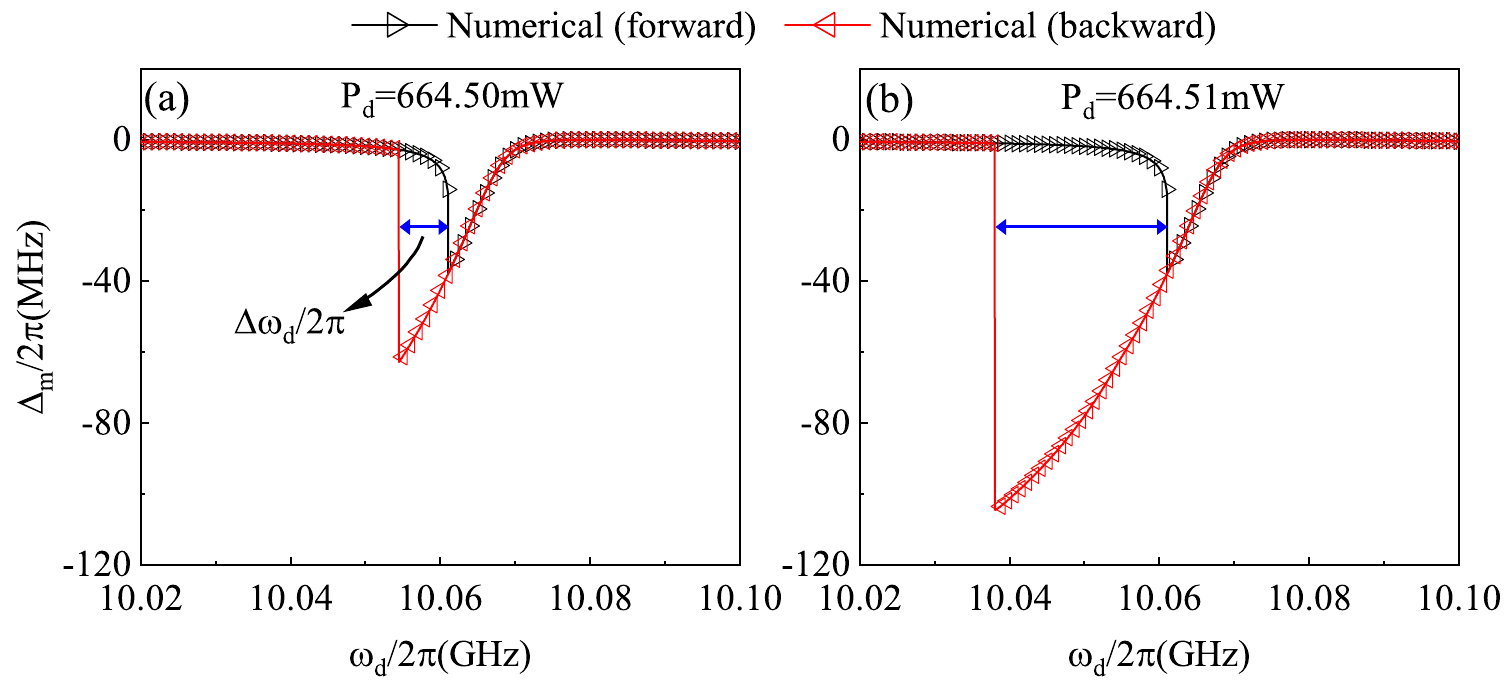}\\
  \caption{Numerically simulated magnon frequency shift $\Delta_{m}/2\pi$ versus the driving frequency $\omega_{d}/2\pi$ for (a) $P_{d}=664.50$ mW and (b) $P_{d}=664.51$ mW when driving the photon-like polariton mode of the lower branch at $\omega_{m}/2\pi=10.18$ GHz. The black and red triangle curves represent the numerically forward and backward sweeps of $\omega_{d}/2\pi$, respectively.}\label{f8}
\end{figure}

Apart from being positive (i.e., $K/2\pi > 0$), the Kerr effect can also be negative (i.e., $K/2\pi < 0$), which can be achieved by adjusting the angle between the external magnetic field and the crystallographic axis. In this case, we can also observe the explosive growth of the bistability when $K/2\pi=-3.5$ GHz, $c/(2\pi)^{3}=-8.0$ MHz$^{3}$/mW, and $\omega_{m}/2\pi=10.18$ GHz, as shown in Fig.~\ref{f8}. In contrast to the positive Kerr effect in Figs.~\ref{f3}(c) and~\ref{f3}(d), the explosive growth of the bistability occurs when driving the photon-like polariton mode of the lower branch [i.e., point $C$ in Fig.~\ref{f1}(b)] instead of the photon-like polariton mode of the upper branch for the negative Kerr effect. Since the reasons for the explosive growth of the bistability are similar in both cases, we will not elaborate them further here.

\section{\label{sec:level4}Discussion and Conclusion}
We have thoroughly analyzed the novel phenomenon of explosive growth of bistability in the cavity magnon system. To provide deeper physical insights, we offer a simple qualitative analysis of the contributions of each term in the Hamiltonian [i.e., Eq.~(\ref{e1})] to the explosive growth of bistability. First, the magnon frequency is influenced by the external magnetic field. The smaller the magnetic field, the lower the magnon frequency and the corresponding energy. In this situation, magnons no longer resonate with microwave photons, leading to a gradual reduction in the magnon component and an increase in the photon component within the upper branch. Since the Kerr nonlinearity resides in the magnon mode, the system enables the photon-like polariton mode to achieve significant bistable behavior through the mechanism of explosive growth of bistability. Second, strong driving is a crucial condition for inducing nonlinear behavior. As the driving power increases and the driving frequency scans toward the photon-like polariton mode, the conditions for explosive growth of bistability become more readily satisfied. In the probe Hamiltonian, the weak probe field does not disturb the explosive growth of bistability but instead reveals this behavior through microwave transmission spectra. Furthermore, in non-conservative systems, the dissipation rate directly affects the energy broadening of both microwave photons and magnons. The energy broadening of the system is positively correlated with the threshold power for explosive growth of bistability. The smaller the energy broadening, the lower the driving power threshold required to achieve explosive growth of bistability.

In this paper, we only consider the uniaxial anisotropy in the YIG sphere. It should be noted that, as a cubic crystal, YIG also exhibits cubic anisotropy~\cite{SP:2009}. Since the magnetization deviates slightly from the external magnetic field during its precession (i.e., $M_{z}/M_{s}$$>0.96$ in our study), the cubic anisotropy energy can be simplified into a form similar to that of the uniaxial anisotropy energy in Eq.~(\ref{e4}). Therefore, the cubic anisotropy does not affect the observed physical phenomena.

In conclusion, we have studied the nonlinear effects for driving different polariton modes in a cavity magnonic system with the magnon Kerr effect. As the driving power increases, the bistability can experience the sudden transition when the photon-like polariton mode is driven, leading to an explosive growth of the bistable region by several times. However, driving the magnon-like polariton mode merely results in the normal bistability. This is due to the fact that, in the former case, the minimum driving power necessary to induce bistability displays a non-monotonic dependency on the driving frequency, whereas in the latter case, it exhibits a monotonic dependency. The critical magnon frequency that triggers the explosive growth of the bistability is regulated by the system parameters. It increases with the enhancement of the photon dissipation rate, but decreases with the enhancement of the coupling strength and the magnon dissipation rate. Furthermore, due to the magnon-photon coupling, the photon- and magnon-like polariton modes can exhibit simultaneous explosive growth of the bistability in microwave transmission when the photon-like polariton mode is driven. Our findings reveal the diversity and complexity of the nonlinear behavior in the cavity magnonic system, which not only deepens our understanding of the nonlinearity in this system but also opens up more possibilities for information processing based on the bistability.

\begin{acknowledgments}
This work was supported by the National Key Research and Development Program of China (Grant No. 2022YFA1405200), the National Natural Science Foundation of China (NNSFC; Grants No. 12204374, No. 12105165, No. 12275212, No. 12174158, and No. 62201268), the Shaanxi Fundamental Science Research Project for Mathematics and Physics (Grant No. 22JSY008), the Technology Innovation Guidance Special Fund of Shaanxi Province (Grant No. 2024QY-SZX-17), and the Youth Innovation Team of Shaanxi Universities (Grant No. 24JP177).
\end{acknowledgments}

\appendix
\renewcommand{\appendixname}{APPENDIX}
\section{APPENDIX: DERIVATION OF EQ.~(\ref{e10}) VIA A MASTER EQUATION METHOD}
The master equation for the density matrix $\rho$ of the system can be written, in the Lindblad form, as~\cite{BP:2007}
\begin{align}\label{ae1}
   \dot{\rho}&=-i\left[H,\rho\right]+\frac{\kappa_{c}}{2}\left( 2a\rho a^{\dag}-a^{\dag}a\rho-\rho a^{\dag}a\right)\nonumber \\
   &\quad +\frac{\gamma_{h}}{2}\sum^{N}_{j=1}\left(2s^{-}_{j}\rho s^{+}_{j}-s^{+}_{j}s^{-}_{j}\rho-\rho s^{+}_{j}s^{-}_{j}\right)\nonumber \\
   &\quad +\frac{\gamma_{m}}{2}\sum^{N}_{j=1}\left(s^{z}_{j}\rho s^{z}_{j}-\frac{\rho}{4}\right)
\end{align}
where $a^{\dag}$ ($a$) is the creation (annihilation) operator of the photon, $\textbf{s}_{j}$=$(s_{j}^{x},s_{j}^{y},s_{j}^{z})$ represents the spin operator of the $j$th spin in the YIG sphere, ${s}_{j}^{\pm}$=$s_{j}^{x}\pm is_{j}^{y}$ are the corresponding raising and lowering operators, $\kappa_{c}$ describes the photon dissipation rate, $\gamma_{h}$ and $\gamma_{m}$ denote the radiative dissipation rate and the nonradiative dephasing rate of the individual spin, respectively. According to the master equation in Eq.~(\ref{ae1}) and the Hamiltonian in Eq.~(\ref{e9}), we can derive the equation of motion for the expectation value of the operator $\mathcal{O}$ (=$\{a, s_{j}^{-}, s_{j}^{z}\}$) via the relation $\langle \dot{\mathcal{O}}\rangle$=Tr$[\dot{\rho}\mathcal{O}]$:
\begin{align}\label{ae2}
\begin{split}
     & \left\langle\dot{a}\right\rangle=-i\left[\left(\omega_{c}-i\frac{\kappa_{c}}{2}\right)\left\langle a\right\rangle+g_{s}\sum_{k=1}^{N}\left\langle s_{k}^{-}\right\rangle+\eta e^{-i\omega_{p} t}\right],\\
     & \left\langle\dot{s_{j}^{-}}\right\rangle=-i \Bigg\{\left(\gamma\mu_{0}H-i\frac{2\gamma_{h}+\gamma_{m}}{4}+2K_{z}\sum_{k=1}^{N}\left\langle s_{k}^{z}\right\rangle\right)\left\langle s^{-}_{j} \right\rangle\\
     & \qquad \quad -\left[\left(K_{x}+K_{y}\right)\sum_{k=1}^{N}\left\langle s_{k}^{-}\right\rangle+2g_{s}\left\langle a\right\rangle +2\Omega_{d}e^{-i\omega_{d} t}\right]\left\langle s^{z}_{j}\right\rangle\Bigg\},\\
     & \left\langle\dot{s_{j}^{z}}\right\rangle=-i \Bigg[-i\gamma_{h}\left(\left\langle s_{j}^{z}\right\rangle +\frac{1}{2}\right)+g_{s}\left(\left\langle a\right\rangle\left\langle s_{j}^{+}\right\rangle-\left\langle a^{\dag}\right\rangle\left\langle s_{j}^{-}\right\rangle \right)\\
     & \qquad \quad +\frac{K_{x}+K_{y}}{2}\left(\left\langle s_{j}^{+}\right\rangle\sum_{k=1}^{N}\left\langle s_{k}^{-}\right\rangle-\left\langle s_{j}^{-}\right\rangle\sum_{k=1}^{N}\left\langle s_{k}^{+}\right\rangle\right)\\
     & \qquad \quad +\Omega_{d}\left(e^{-i\omega_{d} t}\left\langle s_{j}^{+}\right\rangle-e^{i\omega_{d} t}\left\langle s^{-}_{j}\right\rangle\right)\Bigg],
\end{split}
\end{align}
where a mean-field approximation is applied to the two operator terms, i.e., $\langle\mathcal{A}\mathcal{B}\rangle$=$\langle\mathcal{A}\rangle\langle\mathcal{B}\rangle$ ($\mathcal{A}$,$\mathcal{B}$=$\{a, s_{j}^{\pm}, s_{k}^{\pm}, s_{j}^{z}, s_{k}^{z}\}$). Summing Eq.~(\ref{ae2}) over all spins in the YIG sphere, we can obtain
\begin{align}\label{ae3}
\begin{split}
     &\left\langle\dot{a}\right\rangle=-i\left[\left(\omega_{c}-i\frac{\kappa_{c}}{2}\right)\left\langle a\right\rangle+g_{s}\left\langle S^{-}\right\rangle+\eta e^{-i\omega_{p} t}\right],\\
     & \left\langle\dot{S^{-}}\right\rangle=-i \Bigg[\left(\gamma\mu_{0}H-i\frac{\gamma_{m}}{2}\right)\left\langle S^{-} \right\rangle+2K_{m}\left\langle S_{z}\right\rangle\left\langle S^{-} \right\rangle\\
     & \qquad \quad -2g_{s}\left\langle S_{z}\right\rangle\left\langle a\right\rangle -2\Omega_{d}\left\langle S_{z}\right\rangle e^{-i\omega_{d} t}\Bigg],\\
     &\left\langle\dot{S_{z}}\right\rangle=-i \Bigg[-i\frac{\gamma_{m}}{2}\left(\left\langle S_{z}\right\rangle +\frac{N}{2}\right)+g_{s}\left(\left\langle a\right\rangle\left\langle S^{+}\right\rangle-\left\langle a^{\dag}\right\rangle\left\langle S^{-}\right\rangle \right)\\
     & \qquad \quad +\Omega_{d}\left(\left\langle S^{+}\right\rangle e^{-i\omega_{d} t}-\left\langle S^{-}\right\rangle e^{i\omega_{d} t}\right)\Bigg],
\end{split}
\end{align}
where the macrospin operators $S_{z}$=$\sum_{j=1}^{N}s_{j}^{z}$ and $S^{\pm}$=$\sum_{j=1}^{N}s_{j}^{\pm}$ are used. If we set $\gamma_{h}=\gamma_{m}/2$, and $(2K_{z}-K_{x}-K_{y})/2=K_{m}$ (denoting the magnon Kerr coefficient), Eq.~(\ref{ae3}) can be simplified to Eq.~(\ref{e10}) in the main text.

%

\end{document}